\documentstyle[twoside,psfig]{article}

\catcode`\@=11
\long\def\@makefntext#1{
\protect\noindent \hbox to 3.2pt {\hskip-.9pt  
$^{{\eightrm\@thefnmark}}$\hfil}#1\hfill}		%CAN BE USED 

\def\@makefnmark{\hbox to 0pt{$^{\@thefnmark}$\hss}}	%ORIGINAL 
	
\def\ps@myheadings{\let\@mkboth\@gobbletwo
\def\@oddhead{\hbox{}
\rightmark\hfil\eightrm\thepage}   
\def\@oddfoot{}\def\@evenhead{\eightrm\thepage\hfil
\leftmark\hbox{}}\def\@evenfoot{}
\def\sectionmark##1{}\def\subsectionmark##1{}}

\oddsidemargin=\evensidemargin
\newcounter{sectionc}\newcounter{subsectionc}\newcounter{subsubsectionc}
\renewcommand{\section}[1] {\vspace{12pt}\addtocounter{sectionc}{1} 
\setcounter{subsectionc}{0}\setcounter{subsubsectionc}{0}\noindent 
	{\tenbf\thesectionc. #1}\par\vspace{5pt}}
\renewcommand{\subsection}[1] {\vspace{12pt}\addtocounter{subsectionc}{1} 
	\setcounter{subsubsectionc}{0}\noindent 
	{\bf\thesectionc.\thesubsectionc. {\kern1pt \bfit #1}}\par\vspace{5pt}}
\renewcommand{\subsubsection}[1] {\vspace{12pt}\addtocounter{subsubsectionc}{1}
	\noindent{\tenrm\thesectionc.\thesubsectionc.\thesubsubsectionc.
	{\kern1pt \tenit #1}}\par\vspace{5pt}}
\newcommand{\nonumsection}[1] {\vspace{12pt}\noindent{\tenbf #1}
	\par\vspace{5pt}}

\newcounter{appendixc}
\newcounter{subappendixc}[appendixc]
\newcounter{subsubappendixc}[subappendixc]
\renewcommand{\thesubappendixc}{\Alph{appendixc}.\arabic{subappendixc}}
\renewcommand{\thesubsubappendixc}
	{\Alph{appendixc}.\arabic{subappendixc}.\arabic{subsubappendixc}}

\renewcommand{\appendix}[1] {\vspace{12pt}
        \refstepcounter{appendixc}
        \setcounter{figure}{0}
        \setcounter{table}{0}
        \setcounter{lemma}{0}
        \setcounter{theorem}{0}
        \setcounter{corollary}{0}
        \setcounter{definition}{0}
        \setcounter{equation}{0}
        \renewcommand{\thefigure}{\Alph{appendixc}.\arabic{figure}}
        \renewcommand{\thetable}{\Alph{appendixc}.\arabic{table}}
        \renewcommand{\theappendixc}{\Alph{appendixc}}
        \renewcommand{\thelemma}{\Alph{appendixc}.\arabic{lemma}}
        \renewcommand{\thetheorem}{\Alph{appendixc}.\arabic{theorem}}
        \renewcommand{\thedefinition}{\Alph{appendixc}.\arabic{definition}}
        \renewcommand{\thecorollary}{\Alph{appendixc}.\arabic{corollary}}
        \renewcommand{\theequation}{\Alph{appendixc}.\arabic{equation}}
        \noindent{\tenbf Appendix \theappendixc #1}\par\vspace{5pt}}
\newcommand{\subappendix}[1] {\vspace{12pt}
        \refstepcounter{subappendixc}
        \noindent{\bf Appendix \thesubappendixc. {\kern1pt \bfit #1}}
	\par\vspace{5pt}}
\newcommand{\subsubappendix}[1] {\vspace{12pt}
        \refstepcounter{subsubappendixc}
        \noindent{\rm Appendix \thesubsubappendixc. {\kern1pt \tenit #1}}
	\par\vspace{5pt}}

\newcommand{\textlineskip}{\baselineskip=13pt}
\newcommand{\smalllineskip}{\baselineskip=10pt}

\def\eightcirc{
\begin{picture}(0,0)
\put(4.4,1.8){\circle{6.5}}
\end{picture}}
\def\eightcopyright{\eightcirc\kern2.7pt\hbox{\eightrm c}} 

\newcommand{\copyrightheading}[1]
	{\vspace*{-2.5cm}\smalllineskip{\flushleft
	{\footnotesize Modern Physics Letters A, #1}\\
	{\footnotesize $\eightcopyright$\, World Scientific Publishing
	 Company}\\
	 }}

\newcommand{\publisher}[2]{{\begin{center}\footnotesize\smalllineskip 
	Received #1\\
	Revised #2
	\end{center}
	}}

\def\abstracts#1#2#3{{
	\centering{\begin{minipage}{4.5in}\footnotesize\baselineskip=10pt
	\parindent=0pt #1\par 
	\parindent=15pt #2\par
	\parindent=15pt #3
	\end{minipage}}\par}} 

\renewenvironment{thebibliography}[1]
	{\frenchspacing
	 \ninerm\baselineskip=11pt
	 \begin{list}{\arabic{enumi}.}
        {\usecounter{enumi}\setlength{\parsep}{0pt}     
	 \setlength{\leftmargin 12.7pt}{\rightmargin 0pt} %FOR 1--9 ITEMS
         \setlength{\itemsep}{0pt} \settowidth
	{\labelwidth}{#1.}\sloppy}}{\end{list}}

\newcounter{itemlistc}
\newcounter{romanlistc}
\newcounter{alphlistc}
\newcounter{arabiclistc}

\newcommand{\fcaption}[1]{
        \refstepcounter{figure}
        \setbox\@tempboxa = \hbox{\footnotesize Fig.~\thefigure. #1}
        \ifdim \wd\@tempboxa > 5in
           {\begin{center}
        \parbox{5in}{\footnotesize\smalllineskip Fig.~\thefigure. #1}
            \end{center}}
        \else
             {\begin{center}
             {\footnotesize Fig.~\thefigure. #1}
              \end{center}}
        \fi}

\newcommand{\tcaption}[1]{
        \refstepcounter{table}
        \setbox\@tempboxa = \hbox{\footnotesize Table~\thetable. #1}
        \ifdim \wd\@tempboxa > 5in
           {\begin{center}
        \parbox{5in}{\footnotesize\smalllineskip Table~\thetable. #1}
            \end{center}}
        \else
             {\begin{center}
             {\footnotesize Table~\thetable. #1}
              \end{center}}
        \fi}

\def\@citex[#1]#2{\if@filesw\immediate\write\@auxout
	{\string\citation{#2}}\fi
\def\@citea{}\@cite{\@for\@citeb:=#2\do
	{\@citea\def\@citea{,}\@ifundefined
	{b@\@citeb}{{\bf ?}\@warning
	{Citation `\@citeb' on page \thepage \space undefined}}
	{\csname b@\@citeb\endcsname}}}{#1}}

\newif\if@cghi
\def\cite{\@cghitrue\@ifnextchar [{\@tempswatrue
	\@citex}{\@tempswafalse\@citex[]}}
\def\citelow{\@cghifalse\@ifnextchar [{\@tempswatrue
	\@citex}{\@tempswafalse\@citex[]}}
\def\@cite#1#2{{$\null^{#1}$\if@tempswa\typeout
	{IJCGA warning: optional citation argument 
	ignored: `#2'} \fi}}

\def\pmb#1{\setbox0=\hbox{#1}
	\kern-.025em\copy0\kern-\wd0
	\kern.05em\copy0\kern-\wd0
	\kern-.025em\raise.0433em\box0}

\def\fnt#1#2{\footnotetext{\kern-.3em
	{$^{\mbox{\scriptsize #1}}$}{#2}}}

\def\fpage#1{\begingroup
\voffset=.3in
\thispagestyle{empty}\begin{table}[b]\centerline{\footnotesize #1}
	\end{table}\endgroup}

\def\runninghead#1#2{\pagestyle{myheadings}
\markboth{{\protect\footnotesize\it{\quad #1}}\hfill}
{\hfill{\protect\footnotesize\it{#2\quad}}}}
\font\tenrm=cmr10
\font\tenit=cmti10 
\font\tenbf=cmbx10
\font\bfit=cmbxti10 at 10pt
\font\ninerm=cmr9

\font\eightrm=cmr8

\def\qed{\hbox{${\vcenter{\vbox{			%HOLLOW SQUARE
   \hrule height 0.4pt\hbox{\vrule width 0.4pt height 6pt
   \kern5pt\vrule width 0.4pt}\hrule height 0.4pt}}}$}}

\newcommand{\be}{\begin{equation}}
\newcommand{\ee}{\end{equation}}
\newcommand{\br}{\begin{eqnarray}}
\newcommand{\er}{\end{eqnarray}}
\newcommand{\beeq}{\begin{equation}}
\newcommand{\eneq}{\end{equation}}
\newcommand{\beeqar}{\begin{eqnarray}}
\newcommand{\eneqarr}{\end{eqnarray}}
\begin{document}
\setlength{\textheight}{7.7truein}  %for 2nd page onwards

\runninghead{
A No-go theorem for de Sitter Compactifications....
$\ldots$}{
A No-go theorem for de Sitter Compactifications....
$\ldots$}

\normalsize\textlineskip
\thispagestyle{empty}
\setcounter{page}{1}

\copyrightheading{}			%{Vol. 0, No.0 (1992) 000--000}

\vspace*{0.88truein}

\fpage{1}
\centerline{\bf 
A NO-GO THEOREM FOR DE SITTER COMPACTIFICATIONS?}
\baselineskip=13pt
\centerline{\footnotesize 
N.D. HARI DASS.}
%\footnote{Typeset names in
%10 pt Times Roman, uppercase. Use the footnote to indicate the
%present or permanent address of the author.}
\baselineskip=12pt
\centerline{\footnotesize\it 
Institute of Mathematical Sciences, C.I.T Campus,}
\baselineskip=10pt
\centerline{\footnotesize\it 
Chennai 600113 ,
INDIA
%\footnote{State completely without abbreviations, the
%affiliation and mailing address, including country. 
%Typeset in 8 pt Times Italic.}
}
\vspace*{10pt}
\publisher{(received date)}{(revised date)}

\vspace*{0.21truein}
\abstracts{A general framework for studying 
compactifications in supergravity and
string theories was introduced by Candelas, Horowitz,
Strominger and Witten \cite{cand}. This was further
generalised to take into account the warp factor
by de Wit, Smit and Hari Dass \cite{us}.Though the
prime focus of the latter was to find solutions
with nontrivial warp factors (shown not to exist
under a variety of circumstances), it was shown there 
that de Sitter compactifications are generically
disfavoured (see also \cite{others}). In this note 
we place these results in
the context of a revived interest in de Sitter spacetimes
.}{}{}

%\vspace*{10pt}
%\keywords{The contents of the keywords}

%\textlineskip			%) USE THIS MEASUREMENT WHEN THERE IS
%\vspace*{12pt}			%) NO SECTION HEADING

\vspace*{1pt}\textlineskip	%) USE THIS MEASUREMENT WHEN THERE IS
%\section{General Appearance}	%) A SECTION HEADING
%\vskip 1.0cm
%\bigskip 
%\vskip 4.0cm

\section{Introduction and preliminaries.}
There is renewed interest in de Sitter spacetimes
both from the microscopic quantum gravity point of
view \cite{witten} as well as from the macroscopic
cosmological point of view \cite{varun}. This interest
has been triggered on the one hand by cosmological
observations pointing towards an {\em accelerating}
universe and on the other hand by a variety of 
conceptual issues with de Sitter quantum gravity.
Models for explaining an accelerating universe make use
of either a {\em cosmological constant} of the
right sign or {\em exotic} matter(also called
{\em quintessence}) which can provide
{\em negative} pressure. It is also possible to
invoke both. The interesting question
is whether Superstring and Supergravity theories
can naturally accommadate either or both of these.

A lot of progress has been made in our understanding
of supergravity and superstring theories. Nevertheless,
attempts to confront these developments with the
known phenomenology of elementary particles have not
been easy. In a pioneering work, Candelas et 
al \cite{cand} set up a general framework to analyse
the ground state configurations of such theories
which admit compactification of the higher dimensional
theories into a four-dimensional Minkowski spacetime
and an internal manifold usually taken to be compact.
More generally they looked for
compactifications of the type
$$
M^d\rightarrow M^4\times M^{d-4}
$$
where $M^{d-4}$ is a {compact} space and $M^4$
is {maximally symmetric} space-time (Minkowski,
(anti-)de Sitter).
%\be
In Candelas et al \cite{cand} this was done by taking
\be
g_{MN}(x,y)=\left(\begin{array}{cc}
                  g^0_{\mu\nu}(x)~~&~~0\\
		  0~~~~&~~g_{mn}(y)
		  \end{array}\right)
\ee
%\ee
%\be
However, the most general metric (in suitable choice
of coordinates) \cite{wein} is of the form
($\Delta \ge 0$)
\be \label{metric}
g_{MN}(x,y)=\left(\begin{array}{cc}
                  { {\Delta^{-1}(y)}}g^0_{\mu\nu}(x)~~&~~0\\
		  0~~~~&~~g_{mn}(y)
		  \end{array}\right)
\ee
%\ee
%\be
The corresponding vielbein (upto a tangent space
rotation) is
\be \label{viel}
E_M^N(x,y)=\left(\begin{array}{cc}
                  { {\Delta^{-1/2}(y)}}e_{\mu}^{0\alpha}(x)~~&~~0\\
		  0~~~~&~~e_m^a(y)
		  \end{array}\right)
\ee
%\ee
%\be
%\be
We have addressed the issue of the restrictions placed on $M^4$ from
the higher dimensional Einstein's equations on the one hand(sec.2) and 
from requirements of residual supersymmetry on the other(sec.3).
In the case of the former what is really required is to split the higher
dimensional equation into its four-dimensional counterpart as well as the
$d-4$-dimensional counterpart. The technical ingredients required for
this are outlined in Appendix A. An important step in this direction is 
a general categorisation of energy-momentum tensors; all the known
supergravity theories come under this class. It is then shown that the
compactness of the internal manifold imposes severe restrictions and
the most important consequence is that de Sitter compactifications are
ruled out. The analysis of residual supersymmetry is in itself not very 
restrictive but when combined with Bianchi identities they too turn out
to give the same restrictions. Because of the very general nature of our
considerations, it would be very hard to evade our conclusions. In sec.4
we have analysed the extent to which the compactness of $M^{d-4}$ can be
relaxed.

\section{On-shell analysis.}
We start by writing
down the higher-dimensional Einstein equations:
\be
{\hat R}_{MN}-{1\over 2}g_{MN}{\hat R}+{\hat T}_{MN}=0
\ee
%\ee
%\be
equivalently
\be
{\hat R}_{MN}+{\hat T}_{MN}-{1\over d-2}g_{MN}{\hat T}=0
\ee
%\ee
%\be
In these equations ${\hat T}_{MN}$ is the higher-dimensional
stress tensor and ${\hat T}$ the higher-dimensional trace of it. 
On using the maximal symmetry of $M^4$(see eqn(A.8)) one finds
\be
{\hat T}=4t+T
\ee
%\ee
%\be
Decomposing the higher-dimensional Einstein equation into
its 4-dimensional and \\6-dimensional parts:
\be \label{four}
3m_4^2-{1\over 2}\Delta~D^m\left(\Delta^{-3}~\partial_m\Delta\right)+{
\left((d-6)t-T\right)
\over (d-2)\Delta}=0
\ee
%\ee
%\be
\be \label{six}
R_{mn}-2\Delta^{1/2}D_m\left(\Delta^{-3/2}\partial_n\Delta\right)+T_{mn}-g_{mn}{(T+4t)\over d-2}=0
\ee
%\ee
We now consider a class of energy-momentum tensors 
that are {\em quadratic} in {\em p-forms} $F_{M-1..M_p}$
($1\leq p \leq d-1$):
\br \label{energy}
{\hat T}_{MN}&=&\sum_p~{\hat T}^{(p)}_{MN}\nonumber\\
{\hat T}^{(p)}_{MN}&=& p{\hat F}_{MQ_1..Q_{p-1}}{\hat F}_N^{Q_1...Q_{p-1}}
-{g_{MN}\over 2}{\hat F}_{Q_1..Q_p}{\hat F}^{Q_1..Q_p}
\er
Such energy-momentum tensors occur for antisymmetric 
gauge fields (${\hat F}$ is the covariant field 
strength) ($3 \leq p \leq d-3$),Yang-Mills field 
strengths ($p=2,d-2$), and scalar fields ($p=1,d-1$).
All the supergravity theories considered here have
energy mimentum tensors of this type \cite{mth}.
In compactifications with a  {\em maximally symmetric}
4-dimensional spacetime the  
{nonvanishing components} are
\br
{\hat F}_{m_1...m_p}&=&F_{m_1...m_p}\nonumber\\
{\hat F}_{\mu\nu\rho\sigma m_1...m_{p-4}}&=&ie^0\epsilon_{\mu\nu\rho\sigma}f_{m_1...m_{p-4}}~~~~(p\geq 4)
\er
This leads to
\br
T^{(p)}_{mn}&=&pF_{mq_1..q_{p-1}}F_n^{q_1..q_{p-1}}\nonumber\\
& &-~^pC_5~ \Delta^4f_{mq_1..q_{p-5}}f_n^{q_1..q_{p-5}}\nonumber\\
& &-{1\over2}g_{mn}(F^2{ -}~^pC_4~\Delta^4f^2)\nonumber\\
t^{(p)}&=&-{1\over 2}(F^2{ +}~^pC_4~\Delta^4f^2)
\er
so that
\br
(d-6)t^{(p)}-T^{(p)}&=&-(p-1)F^2\nonumber\\
                    & & -^pC_4(d-p+1)\Delta^4f^2
\er
The crucial point is that this is {\em negative} for
$2 \leq p \leq d-2$ and {\em zero} for $p=1,d-1$. Even
though we have considered compactifications of the
type $M^{d-4}\times M^4$ all the remarks generalise
to the $M^{d-D}\times M^D$ cases also.

We now rewrite the four-dimensional field eqn (\ref{four}) as
\be \label{four.2}
{3m_4^2\over \Delta}-{~D^m\left(\Delta^{-3}~\partial_m\Delta\right)\over 2}+{\Delta^{-2}\left((d-6)t-T\right)
\over d-2}=0
\ee
and integrate over the {\em compact} internal space. On
using the {\em negativity} of $(d-6)t-T$ and the fact
that the
integral of the divergence is zero one finds
\be \label{ds.1}
m_4^2\geq 0
\ee
Which means that the {\em de Sitter case 
($m_4^2< 0$) is ruled out!}. If we had started with
 higher dimensional Einstein's equations with a cosmological
 constant $\Lambda$(our conventions are : $\Lambda > 0$
 is {\em de Sitter}), this condition would become
\be \label{ds.2}
3m_4^2+{\Lambda\over (d-2)\Delta}\geq 0
\ee
If the higher dimensional theory is Anti-de Sitter
($\Lambda < 0$), the four dimensional world can only
be {\em anti-de Sitter}.

Returning to the analysis without $\Lambda$, if $M^4$ 
is (flat)Minkowski spacetime
one gets further restrictions
\be \label{mink1}
m_4^2=0~\rightarrow~(d-6)t^{(p)}-T^{(p)}=0
\ee
Consequently
\be
D^m(\Delta^{-3}\partial_m\Delta)=0~~\rightarrow \Delta=const.
\ee
The field strengths are also constrained by eqn(\ref{mink1}):
\be
{\hat F}_{M_1..M_p}=0 ~~~~2\leq p\leq d-2   
\ee
It should be noted that to arrive at eqn (\ref{ds.1})
it is not mandatory that $M^6$ has to be compact; it
suffices for $\Delta(y)$ and the various field 
strengths to have appropriate asymptotic fall off
conditions. This point will be elaborated later.

The derivatives of $\Delta$ do not always occur as in
eqn (\ref{four.2}). In later applications the following remark will be
of importance:
if $f(\Delta)$ is a {\em positive function} and
$a,b$ are {\em arbitrary constants} ($b \ne 0$),
then
\be \label{remark}
f(\Delta)[a(\partial_m\Delta)^2+b\partial_m\partial_m(\ln\Delta)]
=g(\Delta)\partial_m X_m
\ee
where $g(\Delta)$ is another {\em positive
function} 
\be \label{posg}
g(\Delta) = f(\Delta)\Delta^{-a/b}
\ee
and $X_m$ is the vector:
\be \label{vecx}
X_m = b\Delta^{{a-b\over b}}~ \partial_m\Delta 
\ee

%{Proof:}
%
%Let
%$$
%f(\Delta)[a(\partial_m\Delta)^2+b\partial_m\partial_m(\ln\Delta)]
%=g(\Delta)\partial_m X_m
%$$
%with $X_m = h(\Delta)\partial_m\Delta$. Expanding
%both sides this can be rewritten as
%$$
%((a-b){f\over \Delta^2}-gh^\prime)\Delta_{,m}\Delta_{,m}+(b{f\over \Delta}-gh)\Delta_{,m,m}=0
%$$
%which can be solved by
%$$
%gh={bf\over \Delta}~~~~~~gh^\prime=(a-b){f\over \Delta^2}
%$$
%Thus
%$$
%h(\Delta)= b\Delta^{{a-b\over b}}~~~ g(\Delta) = f(\Delta)\Delta^{-a/b}
%$$
\section{Residual Supersymmetry.}

In the previous section we analysed the Einstein's equations to
address the issue of possible compactifications. One may not want 
to impose classical field equations for a variety of reasons. In the
context of string theory, for example, the {\em zero-slope limit}
may yield corrections. Of course, these higher derivative corrections
are very small in the long distance limit of these theories.

Another point to keep in mind is the fact that it is desirable for 
compactifications to maintain some
{\em residual supersymmetry} to control quantum
fluctuations or solve the so called hierarchy
problem.

In what follows we shall use only the {\em On-shell} supersymmetry 
transformations. These along with {\em Bianchi identities} are shown
to imply {\em contracted field equations} though not the {\em full} ones.  
Nevertheless these are quite restrictive.

It is actually desirable to use {\em Off-shell} supersymmetry
transformations. But off-shell formulations of supersymmetry are
very difficult to obtain. They are available only for 
{\em type-I} supergravity theories without super-Yang-Mills multiplets.

In the next few subsections we illustrate how considerations of
residual supersymmetry when used in conjunction with suitable
Bianchi identities turn out to be almost as restrictive as the
on-shell analysis of the previous section. Since only on-shell supersymmetry
transformations are used this will amount to something in between a
fully off-shell analysis and an on-shell analysis. It should be emphasised
that if full supersymmetry is used as versus residual supersymmetry
one would just recover the full bosonic field equations.

\subsection{$d=10, N=1$ Supergravity.}

The field content of this theory, also called 
$d=10, N=1$ Einstein-Maxwell supergravity, is: the 
zehnbein $E_M^A$, the gravitino field
$\Psi_M$, a field strength $H_{MNP}$, a scalar field
$\phi$
and a spinor field $\lambda$. In addition one has
the field strengths $F_{MN}^A$ and
spinors $\chi^A$ for the Yang-Mills multiplet.

The fermionic supersymmetry transformations modulo
terms of higher order in the fermionic fields are:
\br
\delta \Psi_M&=&D_M \epsilon+{\sqrt 2\over 32} e^{2\phi}
(\Gamma_M^{NPQ}-9 \delta_M^N\Gamma^{PQ})H_{NPQ}\epsilon\nonumber\\
\delta\lambda &=& (\sqrt 2 \Gamma^MD_M\phi+{1\over 8}e^{2\phi}\Gamma^{MNP}H_{MNP})\epsilon\nonumber\\
\delta \chi^A&=& -{1\over 4}F^A_{MN}\Gamma_{MN}\epsilon
\er
As we will be working in a background where all fermion
fields vanish, the terms higher order in the fermion
fields do not matter. On using eqn (\ref{viel}) we
can write these as
\br
\delta\psi_\mu&=&D_\mu\epsilon-i\gamma_\mu\gamma_5\Delta^{-1/2}T\epsilon\nonumber\\
\delta\psi_m&=&D_m\epsilon+{\sqrt 2\over 32} e^{2\phi}(\Gamma_mH-12H_m)\epsilon\nonumber\\
\delta\lambda &=&(\sqrt 2i\Gamma^m\partial_m\phi+{1\over 8}ie^{2\phi}H)\epsilon\nonumber\\
\delta\chi^A&=&-{1\over 4}F^A_{mn}\Gamma^{mn}\epsilon
\er
where
\br
T&=&-{1\over 4}i\Gamma^m\partial_m\ln \Delta+{i\sqrt 2\over 32} e^{2\phi}H\nonumber\\
H&=&H^{mnp}\Gamma_{mnp}~~~~H^m=H^{mnp}\Gamma_{np}
\er
The detailed analysis of this theory can be found in
\cite{us}. Here we will only highlight the main line
of reasoning. The {\em integrability condition} for $\delta\psi_\mu=0$
is (after use has been made of the maximal symmetry of
$M^4$):
\be
(T^2+{1\over 4}m_4^2\Delta)\epsilon=0
\ee
Combining this with the $\delta\lambda$ equation one finds
\br \label{lambda}
{1\over 64}\sqrt 2&\{&\Gamma^m\partial_m(\phi+{1\over 2}\ln\Delta),e^{2\phi}H\}\epsilon\nonumber\\
&=&-{1\over 4}\Delta(m_4^2-\partial_m(\Delta^{-1/2}e^{\phi})\partial^m(\Delta^{-1/2}e^{-\phi}))\epsilon
\er
The matrix multiplying $\epsilon$ on the l.h.s is
{\em antihermitian} which can only have {\em imaginary}
eigenvalues. Thus the real eigenvalue occurring in
eqn (\ref{lambda}) must be {\em zero}:
\be \label{typeI}
m_4^2=\partial_m(\Delta^{-1/2}e^\phi)\partial^m(\Delta^{-1/2}e^{-\phi})
\ee
In a compact space {\em every function must have
a maximum} unless it is {\em constant}. Therefore
\be
m_4^2=0~~\rightarrow~\Delta^{1/2}e^\phi=const
\ee
Thus we see that in this case considerations of
residual supersymmetry rule out {\em both de Sitter 
and anti-de Sitter compactifications}. 

For the allowed Minkowski case one can draw further
conclusions. After a lot of algebra (see \cite{us}
for details) one finds
\br \label{1} 
&&{\sqrt 2} e^{2\phi}\Gamma^{mnpq}(D_mH_{npq}+{1\over 4}\sqrt 2 F^A_{mn}F^A_{pq})\eta\nonumber\\
&=&(e^{2\phi}(F^A_{mn})^2+6e^{4\phi}H^2-16D^2\phi-64(\partial_m\phi)^2)\eta
\er
where $\eta$ is a {\em commuting} 6-dimensional spinor.
The {\em classical Bianchi identity} for this theory
is of the form:
\be \label{bianchi}
D_mH_{npq}+{1\over 4}\sqrt 2 F^A_{mn}F^A_{pq}=0
\ee
Using this in eqn (\ref{1}) one gets
\be \label{2}
e^{2\phi}(F^A_{mn})^2+6e^{4\phi}H^2-16D^2\phi-64(\partial_m\phi)^2=0
\ee
Using the result of eqn (\ref{remark}) it is easily
seen that
\be \label{3}
\partial_m\phi=0;~~\partial_m\Delta=0;~~F^A_{mn}=0;~~H_{mnp}=0
\ee
It turns out that {\em anomaly cancellations} in this
theory require the classical Bianchi identity of eqn (\ref{bianchi}) to be modified to
\be \label{qbianchi}
D_{[m}H_{npq]}=-{1\over 4}\sqrt 2 F^A_{[mn}F^A_{pq]}+aR_{[mn}^{ab}R_{pq]}^{ab}
\ee
where $a$ is a constant fixed by the anomaly-free
condition.
Therefore it is no longer possible to obtain eqns (\ref{2},\ref{3}). As was done in \cite{cand}, some components of the 
gauge fields can be identified with
the spin connection
\be \label{gaugespin}
F^A_{[mn}F^A_{pq]}=2\sqrt 2 a R_{[mn}^{ab}R_{pq]}^{ab}
\ee
Quite miraculously an alternate expression emerges out 
of the residual supersymmetry analysis:
\be
{\sqrt 2e^{2\phi}\Gamma DH\over32}\eta
=[{3e^{4\phi}H^2\over 16}-2(\partial_m\phi)^2-
{D^2\phi\over 2}]\eta
\ee
Using this and eqns (\ref{qbianchi},\ref{gaugespin}) one arrives at
\be
{3\over 16}e^{4\phi}H_{mnp}^2-2(\partial_m\phi)^2-{1\over 2}D^mD_m\phi=0
\ee
Applying our earlier techniques to a compact $M^6$ 
one again gets the same conclusions as before. Also 
$M^6$ turns
out to be {\em Ricci-flat}.

\subsection{Nonchiral $d=10,N=2$ Supergravity}
This is a theory that can descend from the 
$d=11,N=1$ supergravity theory. Therefore we analyse
the latter.
The field content of this 
$d=11$ supergravity
theory is 
an elfbein $E_M^A$, a gravitino field $\Psi_M$ and
a field strength $F_{MNPQ}$.
The {\em Bianchi identity} for $F_{MNPQ}$ is
\be
\partial_{[M}F_{PQRS]}=0
\ee

We leave out all the details (which can be found 
in \cite{us}) and just state that the consequences
of residual supersymmetry reduce to the condition
\br
&-&{\sqrt 2\over 432}(\Gamma^{mpqrs}D_mF_{pqrs}+2i\Delta^2\Gamma^m\partial_mf)\epsilon\nonumber\\
&=&\{m_4^2\Delta-{F^2\over 216}-
{2f^2\Delta^4\over(144)^2}+
{(\partial_m\ln\Delta)^2\over 3}-
{D^2\ln\Delta\over 6}\}\epsilon
\er
Again the l.h.s is proportional to the Bianchi identity for $F_{MNPQ}$. Imposing this identity one is led to
\be \label{nchiral}
m_4^2\Delta-{F^2_{mnpq}\over 216}-{2f^2\Delta^4
\over (144)^2}+{(\partial_m\ln\Delta)^2\over 3}
-{D^2\ln\Delta\over 6}=0
\ee
and our earlier techniques applied to a {\em compact}
$M^6$ give the following:
\noindent
{\em The de Sitter case $m_4^2< 0$ is ruled out.}\\
If one considers the flat Minkowski case ($m_4^2=0$)
which is permitted by eqn (\ref{nchiral}) one can
draw further conclusions i.e
\be
m_4^2=0 \rightarrow {\Delta=const,F_{mnpq}=0.}
\ee
Once again the result is that the warp factor has to
be {\em trivial} for the Minkowski case.

The anti-de Sitter case ( $m_4^2> 0$) is also
permitted by eqn (\ref{nchiral}). But in this case
no further restrictions can be extracted. 
Both the warp factor 
and 
the field strengths
($F_{mnpq},f$)
can be {\em nontrivial}. In fact explicit solutions
of this kind are known (for references see \cite{us}).

\subsection{Chiral $d=10,N=2$ Supergravity.} 
This ten dimensional theory(also called IIB) does not 
descend from
$d=11$ supergravity theory.
The field content of this theory consists of a zehnbein $E_M^A$, a gravitino
field $\Psi_M$, a spinor field $\lambda$, a complex
field strength $G_{MNP}$, a self-dual field strength
$F_{MNPQR}$, a complex vector $P_M$ and a $U(1)$-gauge
field $Q_M$.

The selfduality condition for the field strength is:
\be
F_{MNPQR}={i\over 120}E \epsilon_{MNPQRSTUVW}F^{STUVW}
\ee
The Bianchi identities for both $F,G$ take the form:
\br
\partial_{[M}F_{NPQRS]}&=&{5i\over12}G_{[MNP}G^*_{QRS]}\nonumber\\
D_{[M}G_{NPQ]}&=&-P_{[M}G^*_{NPQ]}
\er
There are Bianchi identities for
$Q_M,P_M$ also which can also be viewed as
Maurer-Cartan equations associated with the {coset space}
$SU(1,1)/U(1)$. 
\br \label{maurer}
\partial_{[M}Q_{N]}&=&-iP^*_{[M}P_{N]}\nonumber\\
D_{[M}P_{N]}&=&0
\er
One can solve eqns (\ref{maurer}) to eliminate $P_M,
Q_M$ in terms of two real scalar fields (the dilaton
and axion).
Once again we leave out all the details(which can be
found in \cite{us}) and just state the consequences
of {residual supersymmetry} along with the 
{Bianchi identities:
\be \label{chiral}
3m_4^2\Delta-{D^2\ln\Delta\over 2}-
(\partial_m\ln\Delta)^2-4f^2-{|G_{mnp}|^2\over 48}=0
\ee
Analysing this equation for compact manifolds shows that
{\em de Sitter case is ruled out.}

For the Minkowski case the additional consequences
are
\be
\Delta=const,f_m=0, G_{mnp}=0.
\ee
Further use of the Maurer-Cartan
equations also yields $P_M=0$ and Ricci-flatness.

For the anti-de Sitter case no such 
restrictions are implied by eqn (\ref{chiral}).

\section{Asymptotic fall off in noncompact $M^6$}
The analysis presented so far has treated $M^6$ as a 
{\em compact} space. But with suitable asymptotic
fall-off conditions the same conclusions can be
reached even for {\em noncompact} $M^6$. The volume
of $M^6$ is $\int d^6y \sqrt {det g_{mn}}$. When this 
volume diverges we have the noncompact case. When $M^6$ is 
noncompact it is necessary that the integral $\int d^6y \sqrt{det g_{mn}}
\Delta^{-1}(y)$ exists for an effective four dimensional description
to exist. If we introduce spherical coordinates on $M^6$ and take
{\em spatial infinity} to be at $Y\rightarrow\infty$ where $Y$ is the
radial coordinate, it is clear that $\Delta^{-1}$ must fall off at least
$\simeq Y^{-6}$. A consistency check on this is provided by the
divergence term in eqn (\ref{four.2}). This leads to a surface
integral of the type $\int dS_m \Delta^{-3}\partial_m \Delta$ and
this should vanish asymptotically.The fall-off of $\Delta^{-1}$ is
indeed compatible with this. Now it follows that from eqn (\ref{four.2})
that $t,T$ must fall of {\em faster } than $\Delta$ which is no problem at all.
Under these fall-off conditions the conclusions previously reached on the
basis of on-shell analysis for
the compact case hold in the noncompact case as well.

The generalisation of the analysis based on residual symmetry to the
noncompact case works differently for the three cases considered in 
sec.3. For the type-I case the condition in eqn (\ref{typeI}) was
obtained without recourse to any compactness. In the
noncompact case this equation can have nontrivial solutions.

In order to extend the analysis of the nonchiral case of sec3.2 we
use eqns (\ref{remark},\ref{posg},\ref{vecx}) to cast eqn (\ref{nchiral})
as
\be
m_4^2\Delta-{F^2_{mnpq}\over 216}-{2f^2\Delta^4
\over (144)^2}
-{\Delta^2\over 6}D_m \Delta^{-3}\partial_m \Delta=0
\ee
Dividing the whole equation by $\Delta^2$ we find that extension to 
the noncompact case is exactly along the same lines as that of the
on-shell analysis.

Finally to extend the analysis of sec.3.3 to the noncompact case,
note that by dividing eqn (\ref{chiral}) by $\Delta^2$ one again verifies
that the terms involving the derivatives of $\Delta$ fall off sufficiently
fast and the conclusions reached for the compact case hold again.

\section{Conclusions.}
One sees that our analysis disfavours de Sitter compactifications
in a variety of circumstances. Furthermore, when $M^4$ is flat-Minkowski, it
also shows that the warp factor has to be trivial. The original analysis
assumed that $M^6$ was compact but sec.4 of this note shows how the
analysis can be extended to the noncompact case. 
We had to be content with only the on-shell 
supersymmetry transformations. It is desirable to
find off-shell extensions.

The present
analysis should be extended to cosmological solutions
where only the 3-geometry is taken to be maximally symmetric. There are
indications that the solution space can be much larger in that case 
\cite{cosmo}.
 In addition
to the issue of the cosmological constant one also has to address the issue
of exotic energy momentum tensors i.e quintessence \cite{varun}. Whether
quintessence can occur naturally in these theories is still an
outstanding issue \cite{quint}. 
The implications of
our results for models like the Randall-Sundrum model \cite{sundrum}
should be investigated.

\nonumsection{Acknowledgments}
\noindent
I would like to thank 
Naresh Dadhich and Dharam Ahluwalia 
for
organising a very stimulating meeting.

\newpage
\appendix

\noindent
In this appendix we give various formulae that are important for 
the analysis carried out in the main text.
The anholonomity (Ricci rotation) 
coefficients  for the vielbien of eqn(3) are given by
\be
\Omega_{AB}^C=2E_A^M~E_B^N~\partial_{[M}E_{N]}^C
\ee
%whose {nonzero} components are
%\ee
%\br
%\Omega_{\alpha\beta}^\gamma&=&\Delta^{1/2}\Omega_{\alpha\beta}^{0\gamma}=2\Delta^{1/2}e_\alpha^{0\mu}e_\beta^{0\nu}\partial_{[\mu}e^{0\gamma}_{\nu]}\nonumber\\
%\Omega_{ab}^c&=&2e_a^me_b^n\partial_{[m}e^c_{n]}\nonumber\\
%\Omega_{a\beta}^\gamma&=&-{1\over 2}\delta_\beta^\gamma e_a^m\Delta^{-1}\partial_m\Delta
%\er
%\be
The spin-connection in terms of 
$\Omega_{AB}^C$ is given by :
\be
\omega_M^{AB}={1\over 2}E_M^C\left(\Omega_{AB}^C-\Omega_{BC}^A-\Omega_{CA}^B\right)
\ee
%\ee
%Nonvanishing components of spin connection:
%\br
%\omega_\mu^{\alpha\beta}(x,y)&=&
%\omega_\mu^{0\alpha\beta}(x)=
%{1\over 2}e_\mu^{0\gamma}\left(\Omega_{\alpha\beta}^{0\gamma}-\Omega_{\beta\gamma}^{0\alpha}-\Omega_{\gamma\alpha}^{0\beta}\right)\nonumber\\
%\omega_m^{ab}(x,y)&=&
%\omega_m^{ab}(y)=
%{1\over 2}e_m^c\left(\Omega_{ab}^c-\Omega_{bc}^a-\Omega_{ca}^b\right)\nonumber\\
%\omega_\mu^{ab}(x,y)&=&{1\over 2}e_\mu^{0\alpha}e_b^m(y)\Delta^{-3/2}(y)\partial_m\Delta(y)\nonumber
%\er
%\be
Finally, the higher-dimensional curvature tensor, 
denoted by $\hat R$ is given by:
\be
{\hat R}_{MN}^{AB}=\partial_M\omega_N^{AB}-\partial_N\omega_M^{AB}-2\omega_{[M}^{AC}\omega_{N]}^{CB}
\ee
%\ee
It's nonvanishing components are (for details see \cite{us}):
\br
{\hat R}_{\mu\nu}^{\alpha\beta}&=&R_{\mu\nu}^{\alpha\beta}+{1\over 2}e_\mu^{[0\alpha}e_\nu^{0\beta]}\Delta^{-3}g^{mn}\partial_m\Delta~\partial_n\Delta\nonumber\\
{\hat R}_{\mu n}^{\alpha b}&=&-{1\over 2}e_\mu^{0\alpha}D_n\left(\Delta^{-3/2}e_b^m\partial_m\Delta\right)\nonumber\\
{\hat R}_{mn}^{ab}&=&R_{mn}^{ab}=\partial_m\omega_n^{ab}-\partial_n\omega_m^{ab}-2\omega_{[m}^{ac}\omega_{n]}^{cb}
\er
The maximal symmetry of $M^4$ implies \cite{wein}
\br
R_{\mu\nu}^{\alpha\beta}&=&\partial_\mu\omega_\nu^{\alpha\beta}-\partial_\nu\omega_\mu^{\alpha\beta}-2\omega_{[\mu}^{\alpha\gamma}\omega_{\nu]}^{\gamma\beta}\nonumber\\
&=&2m_4^2e_{[\mu}^{0\alpha}e_{\nu]}^{0\beta}
\er
%\be
We shall also be needing the higher-dimensional Ricci 
tensor:
\be
{\hat R}_{MN}={\hat R}_{MP}^{AB}E_{NA}E_B^P
\ee
%\ee
The nonvanishing components of this Ricci-tensor are:
\br
{\hat R}_{\mu\nu}&=&R_{\mu\nu}-{1\over 2}g^0_{\mu\nu}\Delta D^m\left(\Delta^{-3}\partial_m\Delta\right)\nonumber\\
&=&g^0_{\mu\nu}\{3m_4^2-{1\over 2}\Delta D^m\left(\Delta^{-3}\partial_m\Delta\right)\}\nonumber\\
{\hat R}_{mn}&=&R_{mn}-2\Delta^{1/2}~D_m\left(\Delta^{-3/2}\partial_n\Delta\right)
\er
Maximal symmetry of $M^4$ implies that the components 
of the d-dimensional energy-momentum tensor 
${\hat T}_{MN}$ have the form:
\br
{\hat T}_{\mu\nu}&=&g_{\mu\nu}t=\Delta^{-1}g^0_{\mu\nu}t\nonumber\\
{\hat T}_{mn}&=&T_{mn}~~~~T= g^{mn}T_{mn}
\er
%\vspace*{-0.5pt}
\newpage

\nonumsection{References}

\end{document}